\documentclass{elsart}

\usepackage{natbib}

\usepackage{epsfig}

\usepackage{amssymb}

\begin{document}

\begin{frontmatter}


\title{New Insights into SNR Evolution Revealed by the Discovery of Recombining Plasmas}

\author{Hiroya Yamaguchi}
\address{Harvard-Smithsonian Center for Astrophysics, 60 Garden Street, 
	Cambridge, MA 02138, USA}
\address{RIKEN (The Institute of Physical and Chemical Research), 
	2-1 Hirosawa, Wako, Saitama 351-0198, Japan}
\ead{hyamaguchi@head.cfa.harvard.edu}

\author{Midori Ozawa and Takao Ohnishi}
\address{Kyoto University, Kitashirakawa-oiwake-cho, Sakyo, Kyoto 606-8502, Japan}

\begin{abstract}

We report the discovery of recombining plasmas in three supernova remnants 
(SNRs) with the Suzaku X-ray astronomy satellite. During SNR's evolution, 
the expanding supernova 
ejecta and the ambient matter are compressed and heated by the reverse and 
forward shocks to form an X-ray emitting hot plasma. Since ionization proceeds 
slowly compared to shock heating, most young or middle-aged SNRs have 
ionizing (underionized) plasmas. Owing to high sensitivity of Suzaku, however, 
we have detected radiative recombination continua (RRCs) from the SNRs 
IC~443, W49B, and G359.1--0.5. 
The presence of the strong RRC is the definitive evidence that the plasma is 
recombining (overionized). As a possible origin of the overionization, an interaction 
between the ejecta and dense circumstellar matter is proposed; the highly ionized 
gas was made at the initial phase of the SNR evolution in dense regions, and 
subsequent rapid adiabatic expansion caused sudden cooling of the electrons. 
The analysis on the full X-ray band spectrum of IC~443, which is newly presented 
in this paper, provides a consistent picture with this scenario. 
We also comment on the implications from the fact that all the SNRs 
having recombining plasmas are correlated with the mixed-morphology class. 
\end{abstract}

\begin{keyword}
ISM: individual (IC~443, W49B, G359.1--0.5) \sep Radiation mechanisms: thermal 
\sep Supernova remnants \sep X-rays: ISM

\end{keyword}

\end{frontmatter}

\parindent=0.5 cm

\section{Introduction}
\label{sec:introduction}

Through a supernova (SN) explosion, huge energy (typically of the order of 
$10^{51}$~erg) is released in form of the kinetic energy of the ejecta. 
Since the supernova remnant (SNR) expands supersonically, shock waves 
are formed to compress and heat the ejecta and ambient matter. 
The shocks generally propagate in an extremely low-density ($\lesssim $ 1~cm$^{-3}$) 
interstellar medium (ISM) and hence are nearly collisionless; the shock transition 
occurs on a length scale much shorter than the typical particle mean free path to 
Coulomb collisions. 
Therefore, in the early phases of the SNR evolution, the shocked plasma is expected 
to be far from equilibrium in terms of the particle (electron and ion) temperatures. 
Furthermore, due to the small cross sections for the ionization and recombination 
processes, non-equilibrium ionization (NEI) is expected in most SNRs. 
The ionization timescale, a key parameter for the NEI state, is defined as $n_e t$, 
the product of the electron density 
and the time since the gas was heated. Typically, $n_e t$ is required to be 
$\gtrsim 10^{12}$~cm$^{-3}$~s for collisional ionization equilibrium (CIE) (e.g., Masai 1984). 
It has been reported that most young or middle-aged SNRs in our Galaxy 
(e.g., Tycho: Hwang et al.\ 1998; SN~1006: Yamaguchi et al.\ 2008; 
Cygnus Loop: Miyata et al.\ 2007) or the Large Magellanic Cloud 
(e.g., Hughes et al.\ 1998) exhibit an ionizing (underionized) NEI plasma 
with $n_e t < 10^{12}$~cm$^{-3}$~s, where the rate of the ionization 
process always exceeds the rate of the recombination process.

In contrast, ASCA observations of the SNRs IC~443 and W49B revealed that 
the ionization degrees (or ionization temperature\footnote{The ionization temperature 
is defined to represent the degree of ionization for each individual element.
When the average charge of the atoms is consistent with that in a CIE plasma with 
an electron temperature of $kT_e'$, $kT_z$ is given to equal $kT_e'$.}, $kT_z$) 
of some heavy elements 
were significantly higher than those expected from the electron temperature $kT_e$
(Kawasaki et al.\ 2002; 2005). This suggests that the plasmas are recombining 
(overionized), which is quite unusual for the thermal evolution of SNRs evolving 
in tenuous ISM. However, following XMM-Newton observations showed that 
these SNRs' spectra can be represented by a model of a CIE plasma, 
and the overionization is only marginal (Troja et al.\ 2008; Miceli et al.\ 2006). 
The presence of recombining plasmas in SNRs has, therefore, been a subject of debate.

\section{Recent Suzaku discovery of recombining plasmas}
\label{sec:result}

The discoveries of recombination continua (RRCs) from several SNRs were recent 
breakthroughs made by Suzaku's high sensitivity to extended sources, because 
the presence of the strong RRC is definitive evidence for the recombining plasma. 
Here we review the results of spectroscopic studies of IC~443 (\S\ref{ssec:ic443}), 
W49B (\S\ref{ssec:w49b}), and G359.1--0.5 (\S\ref{ssec:g359}) 
using X-ray Imaging Spectrometer (XIS) on board Suzaku. 
Details for the observations and analysis procedure are found in 
Yamaguchi et al.\ (2009), Ozawa et al.\ (2009), and Ohnishi et al.\ (2011), 
for IC~443, W49B, and G359.1--0.5, respectively.

\subsection{IC~443}
\label{ssec:ic443}

IC~443 (the Jellyfish Nebula) is located near the Gem OB1 association and 
a dense giant molecular cloud (Cornett et al.\ 1977), suggesting that the 
remnant originated from a collapse of a massive star. 
The SNR likely evolved in a low density cavity formed by 
the progenitor's stellar wind activity and recently encountered the cavity wall 
or the molecular cloud (Braun \& Strom 1986; Troja et al.\ 2006; 2008). 
The age of the SNR was estimated to be $\sim$4000~yr from the radii 
of the forward and reverse shocks (Troja et al.\ 2008).

\begin{figure}
\begin{center}
\includegraphics*[width=6cm]{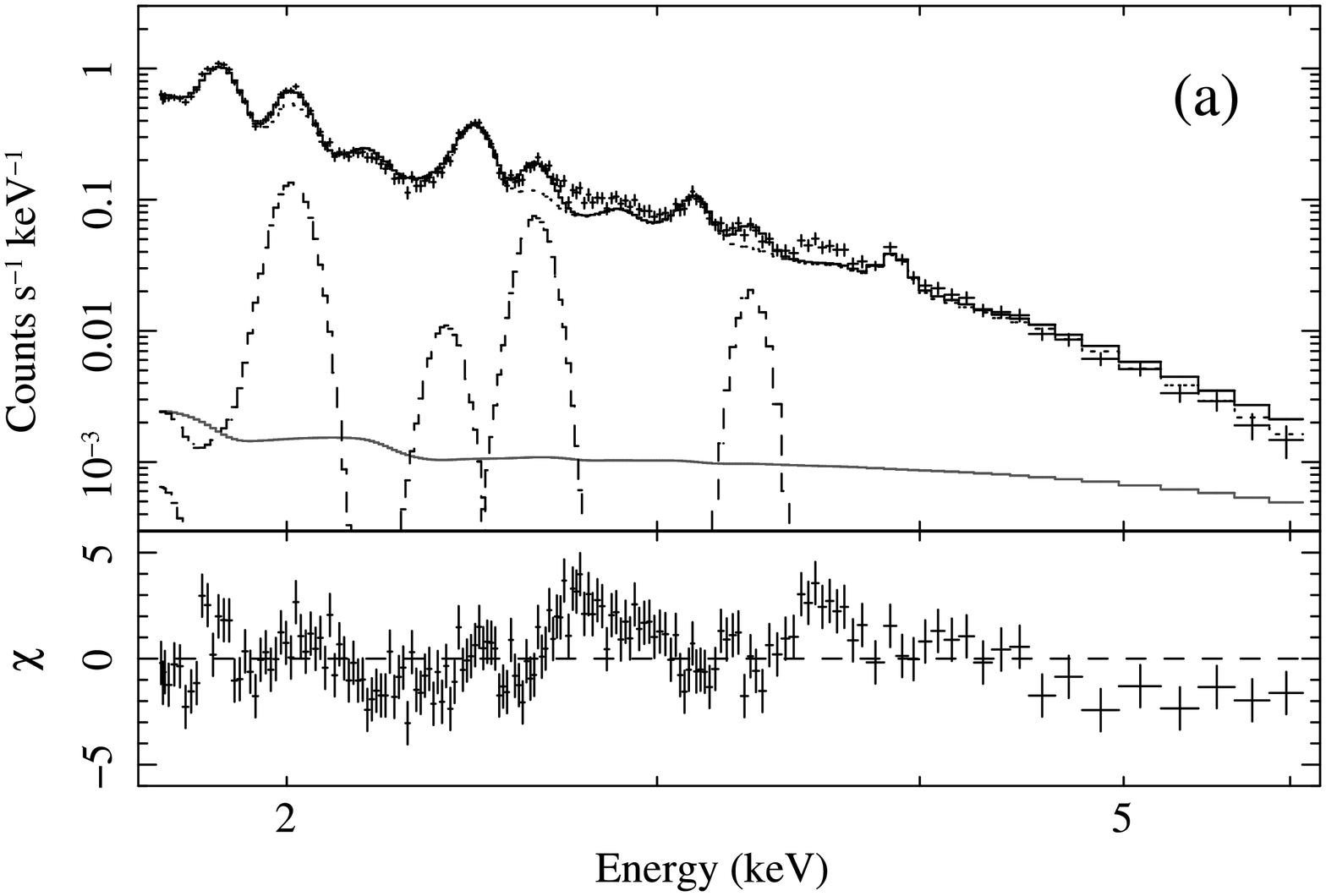}
\includegraphics*[width=6cm]{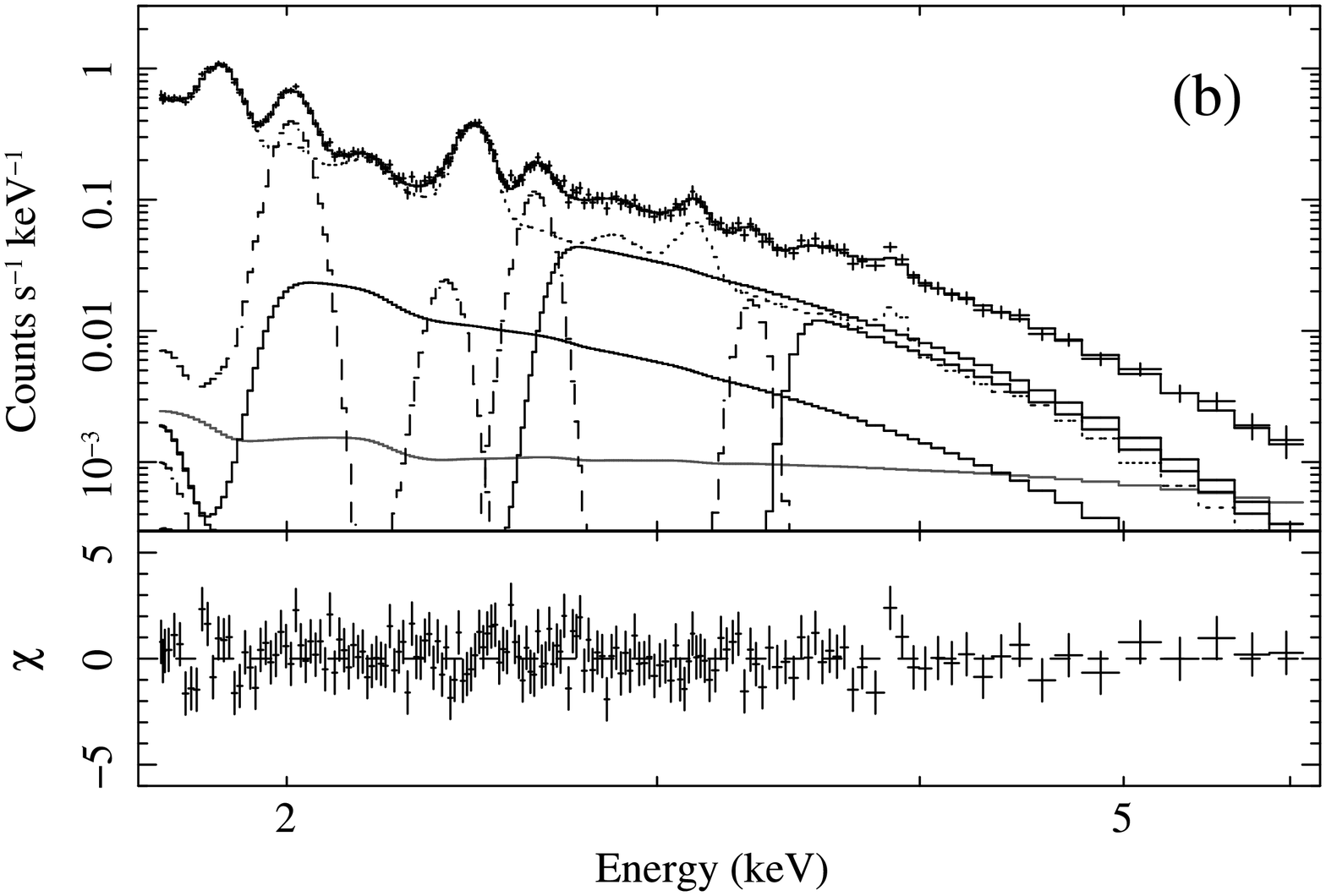}
\end{center}
\caption{(a) XIS spectrum of IC~443 in the 1.75--6.0~keV band, 
fitted with the CIE plasma (black dotted line), additional Lyman lines 
(black dashed lines; see text), and cosmic X-ray background (gray solid line). 
The lower panel shows the residual from the best-fit model.
Two broad features are seen at $\sim$2.7~keV and $\sim$3.5~keV. \ 
(b) Same spectrum as (a), but for a fit with RRC components of H-like Mg, 
Si, and S (black solid lines). The residuals seen in (a) are now gone. 
\label{fig:ic443}
}
\end{figure}

The first claim of overionization in this SNR is presented by the ASCA observation 
(Kawasaki et al.\ 2002). They measured intensity ratios of the H-like to He-like 
K$\alpha$ lines of S to obtain $kT_z$ = 1.5~keV for this element. 
This value is significantly higher than the electron temperature of 1.0~keV 
measured from the slope of the bremsstrahlung continuum spectrum. 
They also found the excess of the intensity in the Si Ly$\alpha$ line over that 
expected from a 1.0~keV CIE plasma, implying the overionization of Si as well. 
However, they were not aware of presence of RRC emissions which are expected 
to be strong in an overionized plasma with the derived temperatures.

Figure~\ref{fig:ic443} shows the XIS spectrum in the energies above 
the Si-K$\alpha$ line extracted from a representative region of the SNR. 
The residuals to a spectral model consisting of a CIE plasma 
(an APEC model: Smith et al.\ 2001) show distinct excess 
Lyman emission from the H-like ions of Si, S, and Ar, as already reported 
by Kawasaki et al.\ (2002). 
After adding Gaussian components to account for these features, 
we found significant hump-like residuals around $\sim$2.7~keV and 
$\sim$3.5~keV (Figure~\ref{fig:ic443}a). These do not correspond to 
emission lines from any abundant element, but are highly consistent with 
the K-shell binding potentials ($I_z$) of the H-like Si (2666~eV) and 
S (3482~eV). Therefore, the residuals likely arise from free-bound transition 
to the K-shells of the H-like Si and S.

When the electron temperature is much lower than the K-edge energy 
($kT_e \ll I_z$), a formula for an RRC spectrum is approximated as 
  \begin{equation}
    \label{eq:1}
 {%
   \frac{dP}{dE}(E_\gamma) \propto 
    {\rm exp}\left( - \frac{E_\gamma - I_z}{kT_e} \right) , ~~ 
    {\rm for}~E_\gamma \geq I_z~.
    }
  \overfullrule 5pt
  \end{equation}
After adding the RRC components, the fit was dramatically improved, 
as shown in Figure~\ref{fig:ic443}b. 
The electron temperature was derived to be $kT_e \sim 0.6$~keV. 
We found that the flux ratios of the H-like RRC to the He-like K$\alpha$ lines 
correspond to the ionization temperatures of $kT_z \sim 1.0$~keV and 
$kT_z \sim 1.2$~keV for Si and S, respectively. 
These values are significantly higher than the electron temperature, 
indicating directly that the plasma is overionized.

\subsection{W49B}
\label{ssec:w49b}

W49B is likely to be associated with the star forming region W49A 
(Brogan \& Troland 2001). The SNR exhibits centrally filled X-rays 
with a bright radio shell. The near-infrared image shows a barrel-shaped 
structure with coaxial rings, which is suggestive of bipolar stellar wind 
structures surrounding a massive progenitor (Keohane et al.\ 2007). 
Therefore, W49B had likely exploded inside a wind-blown bubble. 
The age of the SNR is uncertain, but is roughly estimated to be 
$\sim$1000~yr (e.g., Lopez et al.\ 2009).

Using ASCA data, Kawasaki et al.\ (2005) derived the ionization temperatures 
of Ar and Ca to be $kT_z \sim 2.5$~keV from intensity ratios of the H-like to 
He-like K$\alpha$ emissions. However, a similar diagnostics was not 
performed for Fe, possibly due to the limited photon statistics. 
It should also be noted that emission lines of Cr and Mn were detected in 
the ASCA spectrum (Hwang et al.\ 2000). 
This is the first detection of these elements in a celestial X-ray source.

\begin{figure}
\begin{center}
\includegraphics*[width=6cm]{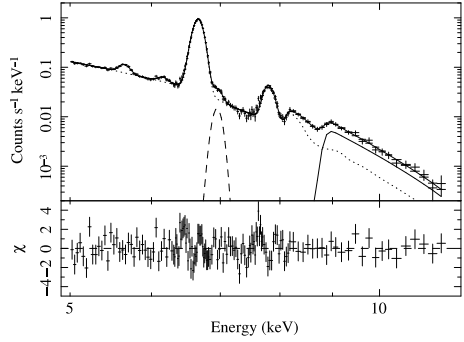}
\end{center}
\caption{XIS spectrum of W49B in the 5.0--12~keV band. 
	The solid, dotted, and dashed lines represent the RRC of He-like Fe,
	CIE plasma, and additional Gaussian accounting for the strong Ly$\alpha$ 
	line, respectively. 
\label{fig:w49b}
}
\end{figure}

Figure~\ref{fig:w49b} shows the XIS spectrum above 5~keV extracted from 
the entire SNR. The spectrum exhibits K$\alpha$ emission lines of 
Cr, Mn, Fe, and Ni (blended with Fe-K$\beta$). 
In addition, we can see a spectral hump at $\sim$9~keV. 
This structure cannot be explained by any combination of high 
temperature plasmas either in the CIE or underionized NEI state, 
but is likely to be caused by an RRC of He-like Fe ($I_z$ = 8830~eV). 
In fact, addition of the RRC, together with the Fe-Ly$\alpha$ line, 
to the CIE plasma component greatly improved the spectral fit. 
The electron and ionization temperatures we derived are 
$kT_e \sim 1.5$~keV, and $kT_z \sim 2.7$~keV, respectively. 
Although we did not detect RRC features of lower-$Z$ elements, 
we confirmed that the H-like to He-like line ratios of Ar and Ca are 
significantly larger than those expected for a 1.5~keV CIE plasma, 
as was firstly claimed by Kawasaki et al.\ (2005). This result will be 
reported in detail in a future paper.

Following our result, Miceli et al.\ (2010) confirmed the presence of 
the strong RRC using XMM-Newton data. Utilizing the higher angular 
resolution, they also investigated the spatial distribution of the recombining 
plasma and found that the highly overionized matters are localized at 
the center of the SNR.

\subsection{G359.1--0.5}
\label{ssec:g359}

G359.1--0.5, an SNR located in the direction of the Galactic center, was discovered by 
a 4.9~GHz radio observation (Downes et al.\ 1979). The angular size of $\sim$$20'$
corresponds to the diameter of $\sim$50~pc (at a distance of 8.5~kpc). 
This large value suggests that G359.1--0.5 is more evolved than IC~443 and W49B. 
The SNR is likely to be interacting with surrounding molecular clouds, 
since bright emissions of 1720~MHz OH maser and 
shocked molecular hydrogen had been detected at the SNR's limb 
(Yusef-Zadeh et al.\ 1995; Lazendic et al.\ 2002).

Using ASCA, Bamba et al.\ (2000) revealed that the X-ray spectrum exhibits 
prominent emission lines of Si and S. They also claimed that the centroid energies 
of the Si- and S-K$\alpha$ lines are consistent with those of He-like Si and 
H-like S (Ly$\alpha$), respectively. This is, if true, quite peculiar because 
the heavier elements tend to be less ionized compared to the lighter ones 
in a normal isothermal plasma. In fact, the model they used to reproduce 
the observed spectrum was composed of two rather non-physical plasma 
components: a cooler one consisting only of Si, and a hotter one with 
extremely over abundant S. 
In the ASCA spectrum, however, the He-like and H-like lines were not 
distinctly separated due to the limited photon statistics and 
the degraded energy resolution.

\begin{figure}
\begin{center}
\includegraphics*[width=6cm]{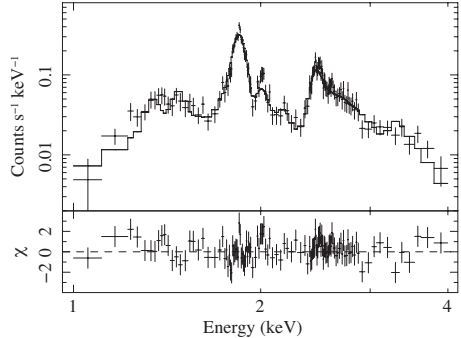}
\end{center}
\caption{XIS spectrum of G359.1--0.5, fitted with an overionized plasma model.
\label{fig:g359}
}
\end{figure}

The recent Suzaku observation has provided an interesting result which 
unveiled the nature of this peculiar spectrum. 
The XIS spectrum of the entire SNR is shown in Figure~\ref{fig:g359}. 
In addition to the K$\alpha$ emission lines of He-like Si/S and H-like Si, 
we can see a broad shoulder-like structure above the energy of the He-like 
S-K$\alpha$ line ($\sim$2.4~keV). 
This structure is analogous to the RRC of Si observed in IC~443. 
We thus introduced a recombining plasma model provided by the SPEX software 
package (Kaastra et al.\ 1996). The best-fit electron and ionization temperatures were 
obtained to be $kT_e \sim 0.3$~keV and $kT_z \sim 0.8$~keV, respectively. 
The recombining state is, therefore, clearly indicated. The shoulder-like spectral 
feature was confirmed to be the RRC of He-like Si ($I_z$ = 2439~eV). 
In the previous study by Bamba et al.\ (2000), 
this feature was misinterpreted as a S-Ly$\alpha$ line and hence presence 
of an unreasonably-high-temperature component was claimed. 
We also attempted to fit the spectrum with a two-component 
CIE plasma model as was performed by Bamba et al.\ (2000), 
but the model failed to reproduce the overall spectrum.

\section{Full X-ray band spectrum of IC~443}
\label{sec:full}

\begin{figure}
\begin{center}
\includegraphics*[width=4.5cm]{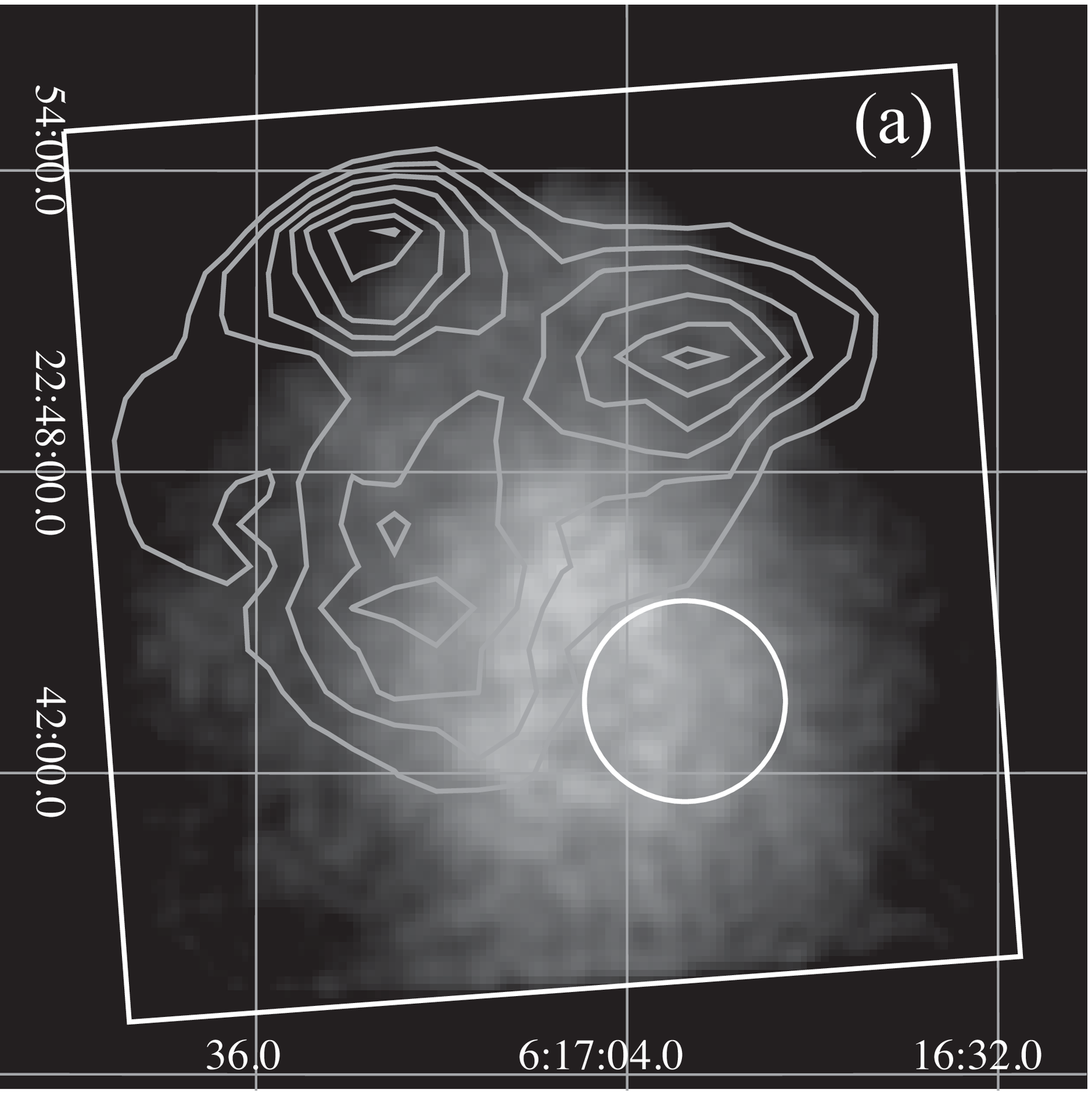}
\includegraphics*[width=6cm]{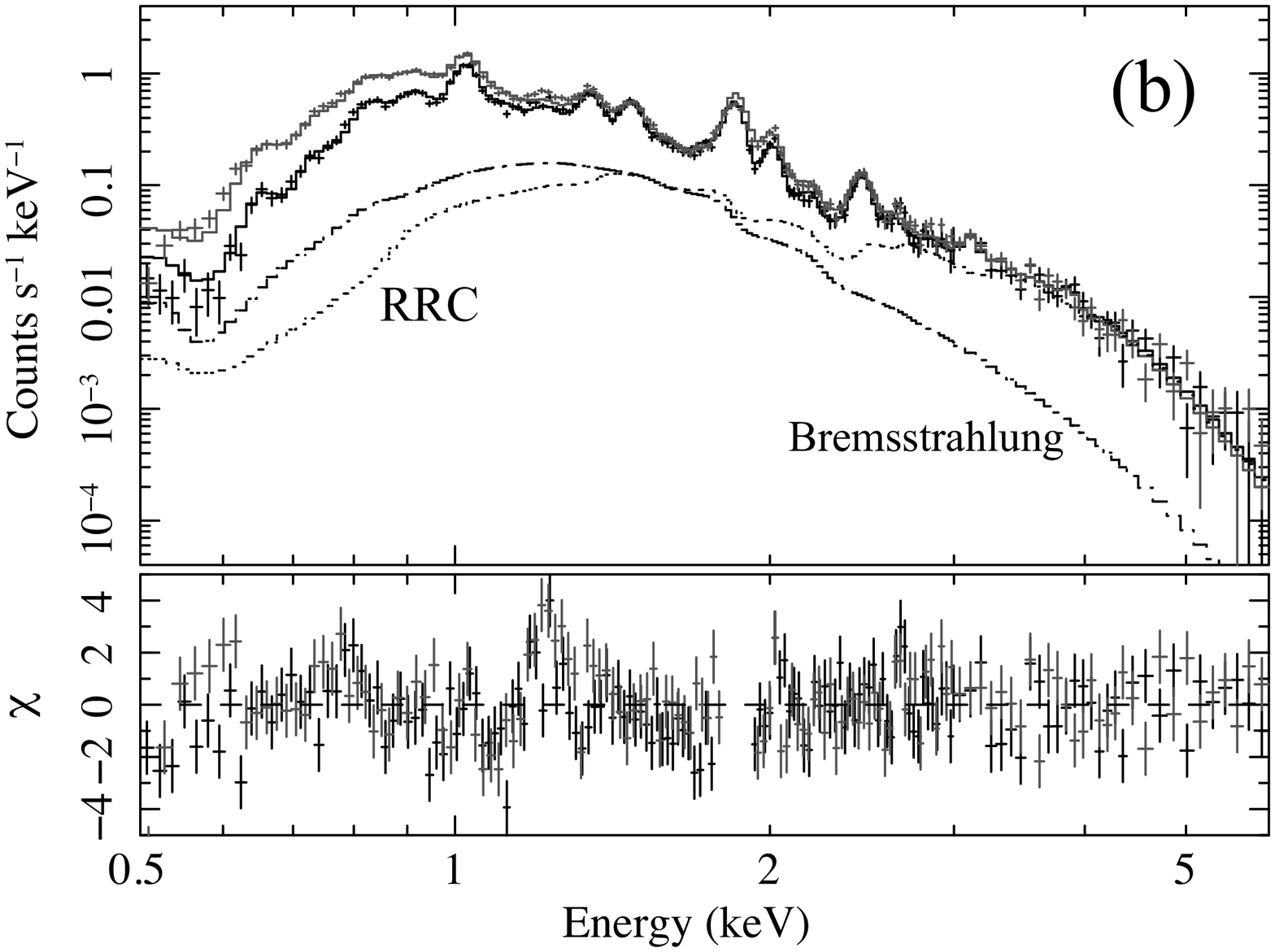}
\end{center}
\caption{(a) XIS image of the northern part of IC~443 in the 1.7--3.0~keV band 
(gray scale), overlaid with contours of the 0.6--1.0~keV emission. 
The coordinates of (R.A.\ and decl.) refer to epoch J2000.0. 
The white square and circle are the XIS field of view and the region where 
the spectrum is extracted, respectively. \ 
(b) Full-band XIS spectrum of IC~443, where the background is subtracted. 
Black and gray represent the data from the XIS0 and XIS1, respectively. 
(The XIS3 data is also analyzed simultaneously, but is not shown for simplicity.) 
The contributions from the bremsstrahlung and RRCs (for the XIS0) are individually indicated. 
\label{fig:ic443_full}
}
\end{figure}

In this section, we present our new result of the spectral analysis on IC~443 
using a recombining plasma model. The analyzed energy range is 
extended to include the soft X-ray band as well. 
Figure~\ref{fig:ic443_full}a shows the 1.7--3.0~keV (hard) image of the northern 
part of IC~443 (Yamaguchi et al.\ 2009), overlaid with contours of the 0.6--1.0~keV 
(soft) band. The shell-like structure seen in the soft X-ray image originates mainly from 
the surrounding clouds which are recently shocked by the blast wave 
(Kawasaki et al.\ 2002; Troja et al.\ 2006). 
To avoid contamination from this emission component as far as possible, we extract 
XIS spectra from a narrow circular region with a radius of 2~arcmin shown in 
Figure~\ref{fig:ic443_full}a. The background was taken from the ``ANTI GALACTIC" 
field (Masui et al.\ 2009) whose aim point was ($l$, $b$) = (235.0, 0.0). 
To minimize the uncertainty due to the background subtraction, in particular that of 
the particle component (non X-ray background), we extracted the background spectra 
from the same detector coordinates as the source regions. 
The background-subtracted spectra are shown in Figure~\ref{fig:ic443_full}b.

As mentioned in Section~\ref{ssec:ic443}, the ionization temperatures of Si and S 
are different from each other (1.0~keV and 1.2~keV, respectively). This implies that 
the single-$kT_z$ model, which was applied for G359.1--0.5, does not work in this case. 
Therefore, we used an NEI model of SPEX, where the free parameters are the initial 
plasma temperature ($kT_{\rm [init]}$), post-shock electron temperature ($kT_e$), 
ionization timescale ($n_e t$), emission measure ($n_e n_{\rm H} V$), 
abundances of different elements, and absorption column ($N_{\rm H}$). 
The first parameter is usually set to a very low value ($\sim$1~eV, for instance) 
to describe a standard underionized NEI plasma, but in our case it should be 
allowed to vary freely and also to be higher than the $kT_e$ value for reproducing 
an overionized plasma.\footnote{More detailed information about this model is found 
at the URL $<$http://www.sron.nl/files/HEA/SPEX/manuals/manual.pdf$>$}
Since the gain calibration of the XIS is reported to be problematic around the energy of 
the Si K-edge (1.84~keV), we ignore the energy range of 1.8--1.9~keV 
from the fitting. The best-fit parameters obtained with this model is given in 
Table~\ref{ic443_fit}. Although the fit is unacceptable from a statistical point of view 
($\chi ^2$/dof = 722/430),\footnote{The largest discrepancy between the data and 
model is seen around 1.2~keV, where the Fe L-shell emissivity is known to be
highly uncertain (e.g., Kosenko et al.\ 2008).} the overall spectrum 
was well reproduced by the model consisting only of one plasma component. 
It is worth noting that an X-ray spectrum from similarly narrow regions in IC~443 
was modeled with at least two components of CIE and/or underionized plasmas with 
different electron temperatures in all the previous works (e.g., Troja et al.\ 2008; 
Bocchino et al.\ 2009). Introducing the recombining plasma model, however, we found 
that multiple temperatures are not always needed. This is because the strong RRCs 
contribute largely to the continuum flux in the hard X-ray band, even though the 
slope of the bremsstrahlung emission is steep due to the low ($\sim$0.6~keV) 
electron temperature (see Figure~\ref{fig:ic443_full}b).

\begin{table}
\caption{Best-fit spectral parameters for IC~443.}
\begin{center}
\begin{tabular}{lc}
\hline
Parameters & Values \\
\hline
$N_{\rm H}$~($10^{21}$~cm$^{-2}$)  &  $5.2 \pm 0.1$ \\
$kT_{\rm [init]}$~(keV)  &  $2.8_{-0.5}^{+0.8}$ \\
$kT_e$~(keV)  &  $0.58 \pm 0.07$ \\
$n_et$~($10^{11}$~cm$^{-3}$~s)  &  $6.7 \pm 0.5$ \\
Ne, Na~(solar)  &  $0.83 \pm 0.04$ \\
Mg, Al~(solar)  &  $0.69_{-0.04}^{+0.05}$ \\
Si~(solar)  &  $1.2 \pm 0.1$ \\
S~(solar)  &  $1.1 \pm 0.1$ \\
Ar, Ca~(solar)  &  $1.3 \pm 0.3$ \\
Fe, Ni~(solar)  &  $0.30 \pm 0.01$ \\
$n_e n_{\rm H} V$~($10^{56}$~cm$^{-3}$)  &  $8.2 \pm 0.3$ \\
\hline
$\chi ^2$/dof & 722/430 \\
\hline
\multicolumn{2}{l}{Statistic errors are for 1$\sigma$ confidence.}
\end{tabular}
\end{center}
\label{ic443_fit}
\end{table}

Assuming the plasma depth of 3~pc, the emitting volume is estimated to be 
$2.2 \times 10^{56}$~cm$^3$. The emission measure is, therefore, converted 
to the uniform electron density of $2.0$~cm$^{-3}$. 
The enhanced abundance ratios of Si/Fe and S/Fe with respect to the solar values are 
typical for core-collapse SNRs, but the Fe abundance would be relatively uncertain 
due to the systematic uncertainty in the Fe L-shell emissivity.

\section{Discussion}
\label{sec:discussion}

\subsection{Origin of the recombining plasmas}
\label{ssec:origin}

Utilizing the high spectral sensitivity of Suzaku, we have discovered 
recombining (overionized) plasmas in the three middle-aged SNRs, 
IC~443, W49B, and G359.1--0.5, for the first time. 
To date, astronomical recombining plasmas had been observed in 
relatively compact regions, such as X-ray binaries 
(e.g., Cygnus~X-3: Kawashima \& Kitamoto 1996) or planetary nebulae 
(e.g., BD+30$^{\circ}$3639: Nordon et al.\ 2009), but those in extended 
sources had not been reported.  As argued in Section~\ref{sec:introduction}, 
SNRs usually have ionizing plasmas as the shock-heated plasma 
slowly achieves CIE. Therefore, the discoveries of the recombining plasmas are 
dramatically changing our understanding of SNR dynamics and evolution.

Recombining plasma can be produced by strong radiation from an external X-ray source 
(photoionization). In IC~443, there is an X-ray pulsar at the center of the remnant 
(CXOU~J061705.3+222127: Olbert et al.\ 2001). However, the X-ray luminosity of this 
source is too faint to produce the ionization balance we observed, as already discussed 
by Kawasaki et al.\ (2002). No central source has been found in W49B and G359.1--0.5. 
Early photoionization by X-ray bursts (i.e., X-ray flash and/or Gamma-ray burst afterglow)
occurred immediately after the SN explosion is also unlikely, because this process 
affects only surrounding ISM and hence cannot explain our results; the SN ejecta 
are highly overionized (Ozawa et al.\ 2009; Ohnishi et al.\ 2011).

\begin{table}
\caption{Summary of the observed parameters and estimations of 
timescale to reach CIE.}
\begin{center}
\begin{tabular}{lccccc}
\hline
~ & $kT_e$ & $n_e$ & Overionized & $t_{\rm CIE}$ & Age  \\
~ & (keV) & (cm$^{-3}$) & elements & (yr) & (yr)  \\
\hline
IC443 & 0.6 & 2.0 & Si, S & $\sim 2 \times 10^4$ & $\sim 4 \times 10^{3\ast}$ \\
W49B & 1.5 & 2.0 & Fe, (Ar, Ca)$^{\ddagger}$ & $\sim 1 \times 10^4$ & 
$\sim 1 \times 10^{3\dagger}$ \\
G359.1--0.5 & 0.3 & 0.2 & Si, S & $\sim 1 \times 10^4$ & Unknown \\
\hline
\multicolumn{6}{l}{$^{\ast}$Troja et al.\ (2008);\ $^{\dagger}$Lopez et al.\ (2009);\ 
$^{\ddagger}$Claimed by Kawasaki et al.\ (2005).}
\end{tabular}
\end{center}
\label{table}
\end{table}

As the mechanism to form the recombining plasma, we here propose rapid adiabatic 
cooling occurred due to fast expansion of SNRs as discussed by Itoh \& Masai (1989). 
If an SN explodes in dense circumstellar matter (CSM) formed by the progenitor's stellar 
wind activity, both CSM and SN ejecta can be shock-heated to high temperature 
at the initial phase of the SNR's evolution. Furthermore, the plasmas can be in almost 
CIE as they have large electron densities. 
Such an CSM/ejecta interaction and a resultant high-temperature CIE plasma 
were observed in SN~1993J at a few days after the explosion 
(Kohmura et al.\ 1994; Uno et al.\ 2002). 
Although the shock velocity is significantly decreased by the dense matter, 
it can quickly recover when the blast wave penetrates into the outer ISM region 
(Itoh \& Masai 1989; a similar suggestion is given by Dwarkadas (2005), 
but for shocks passing through more expanded wind-blown shells). 
If the shocked electrons cool adiabatically much faster than the recombination 
timescale of the ionized materials, overionized plasma can be formed and 
survive for a long time.

This picture is supported by our new result on IC~443 where the overionized NEI 
model is applied. The best-fit parameters we obtained (Table~\ref{ic443_fit}) can be 
interpreted with the scenario that the 2.8~keV ($kT_{\rm [init]}$) plasma in almost CIE 
suddenly cooled down to 0.6~keV ($kT_e$), in the initial evolution phase, to form the 
recombining state. From the ionization timescale of $6.7 \times 10^{11}$~cm$^{-3}$~s, 
we estimate the time since the sudden cooling to be 
$t$ = $1.1 \times 10^4$ $(n_e/2.0~{\rm cm}^{-3})^{-1}$~yr. 
Although the value is about three times larger than the known age of the remnant 
(4000~yr; Troja et al.\ 2008), this is not surprising because the actual evolution must 
be more complex; both the electron density and temperature are time-dependent. 
Future works taking into account the realistic evolutions are, therefore,  
highly encouraged to examine this scenario more precisely.

It should also be noted that an alternative scenario to create the recombining 
plasma has recently been proposed by Zhou et al.\ (2011) as an application 
of hydrodynamic simulation to W49B. They found that adiabatic expansion 
causes rapid cooling of the ejecta (as was argued by Itoh \& Masai 1989) 
but thermal conduction occurring in a mixture of ambient clouds and 
a hot plasma plays an important role to form the recombining plasma. 
This scenario successfully explains the observational result from 
XMM-Newton (Miceli et al.\ 2010); 
the strong overionization is localized at the SNR's center whereas no overionized 
stuff is present in the eastern region where the ejecta expansion is hampered 
by the molecular cloud.

In Table~\ref{table}, we summarize the physical parameters of the three SNRs 
with estimates of the recombination timescales (to reach 90\% of CIE)
using the calculation by Smith \& Hughes (2010). In IC~443 and W49B, 
the timescales are significantly larger than the ages of the remnants. 
Therefore, both the scenarios of the rapid adiabatic expansion and thermal 
conduction are supported. The age of G359.1--0.5 is currently unknown, but 
it should be less than about $1 \times 10^4$~yr if the former scenario is the case.

\subsection{Relation with mixed-morphology class}
\label{ssec:mmsnr}

It is notable that all the three SNRs that have shown the recombining plasmas 
are classified as mixed-morphology (MM) SNRs (Rho \& Petre 1998). 
The MM-SNRs are characterized by a centrally-peaked X-ray profile with 
a more limb-brightened radio morphology. The X-ray emission is thermal 
in nature (in contrast to Crab-like pulsar wind nebulae), and generally exhibits 
a relatively uniform radial temperature profile (e.g., Slane et al.\ 2002). 
These properties are in distinct contrast to the expected properties of evolved 
SNRs in a homogeneous ISM; the Sedov solution predicts a limb-brightened 
X-ray morphology and a temperature profile declining steeply from the SNR 
center to the limb. 
Therefore, more complex theoretical models invoking thermal conduction 
(e.g., Shelton et al.\ 1999) or cloud evaporation (e.g., White \& Long 1991) 
have been proposed to explain a mechanism responsible for the observed 
properties of MM-SNRs. However, application of these models to individual 
SNRs has been largely unsuccessful to date (e.g., Slane et al.\ 2002; 
Bocchino et al.\ 2009), suggesting that the situation is more complex.

So far, all the theoretical models for MM-SNRs have assumed that the plasmas 
are always in CIE. Our new results, however, require a modified picture 
and strongly constrain any model that can provide a significant amount 
of overionized materials. As already mentioned, Itoh \& Masai (1989) predicted 
that the recombining plasma can be created if an SNR initially evolves in 
dense CSM produced by progenitor's stellar winds. 
Interestingly, it is suggested that density gradients in pre-existing CSM could 
be a major contributor to the emission profiles in the MM-SNRs (Petruk 2001). 
All the SNRs we have presented are associated with the massive star forming regions 
(and thus possibly with Type~II progenitors), supporting the presence in the past of 
stellar wind materials. Future efforts for modeling the MM-SNRs, therefore, should 
properly consider the effects of ISM nonuniformity to both the dynamical and ionization 
evolutions in the SNRs. This will help improve our knowledge about the poorly 
understood MM-SNRs as well as unusual overionized plasmas in this class.

\section{Conclusion}

We have discovered the strong RRC emissions from the three MM-SNRs, 
IC~443, W49B, and G359.1--0.5, for the first time. 
Since plasmas heated by forward or reverse shocks are expected to be 
ionizing (underionized) in usual SNRs, this discovery is dramatically 
changing our understanding of SNR dynamics and evolution. 
We have also demonstrated that a one-component overionized plasma 
model can reasonably reproduce the full X-ray band spectrum of IC~443 
that has so far been modeled with at least two components of CIE and/or 
underionized NEI plasmas. 
As the origin of the overionization, rapid adiabatic cooling after the interaction 
between the pre-exiting CSM and SN ejecta is proposed. 
Alternatively, thermal conduction may work effectively to form the recombining 
plasmas in some cases. 
Future theoretical works properly taking into account the ISM nonuniformity 
and ionization nonequilibrium will help reveal the detailed mechanisms 
to form the overionization and centrally-filled X-ray morphology in these 
remnants.

\bigskip 

The authors appreciate a number of constructive suggestions from the anonymous referees. 
We also thank to Drs.\ Katsuji Koyama, Kuniaki Masai, Randall K.\ Smith, Patrick O.\ Slane, 
and Jelle Kaastra for helpful discussion. H.Y.\ is supported by Japan Society for the Promotion 
of Science (JSPS) Research Fellowship for Research Abroad.

\end{document}